# Polarization-controlled volatile ferroelectric and capacitive switching in $Sn_2P_2S_6$

Sabine M. Neumayer[1*], Anton V. Ievlev[1], Alexander Tselev[2], Sergey A. Basun[3,4], Benjamin S. Conner[5,6], Michael A. Susner[3], Petro Maksymovych[1*]

[1]Center for Nanophase Materials Sciences, Oak Ridge National Laboratory, Oak Ridge, Tennessee 37831, United States

[2]CICECO-Aveiro Institute of Materials and Department of Physics, University of Aveiro, Aveiro 3810-193, Portugal

[3]Materials and Manufacturing Directorate, Air Force Research Laboratory, 2179 12th Street, Wright-Patterson Air Force Base, Ohio 45433, USA

[4]Azimuth Corporation, 4027 Colonel Glenn Highway, Suite 230, Beavercreek, Ohio 45431, United States

[5]Sensors Directorate, Air Force Research Laboratory, 2241 Avionics Circle, Wright-Patterson Air Force Base, Ohio 45433, USA

[6]National Research Council, Washington, D.C. 20001, USA

*Corresponding authors:
Petro Maksymovych maksymovychp@ornl.gov
Sabine M. Neumayer neumayersm@ornl.gov

**KEYWORDS**

Domain walls, neuromorphic, tunable, ferroelectric, scanning microwave impedance microscopy

**ABSTRACT**

*Smart electronic circuits that support neuromorphic computing on the hardware level necessitate materials with memristive, memcapacitive, and neuromorphic- like functional properties; in short, the electronic response must depend on the voltage history, thus enabling learning algorithms. Here we demonstrate volatile ferroelectric switching of $Sn_2P_2S_6$ at room temperature and see that initial polarization orientation strongly determines the properties of polarization switching. In particular, polarization switching hysteresis is strongly imprinted by the original polarization state, shifting the regions of non-linearity toward zero-bias. As a corollary, polarization switching also enables effective capacitive switching, approaching the sought-after regime of memcapacitance. Landau-Ginzburg-Devonshire simulations demonstrate that one mechanism by which polarization can control the shape of the hysteresis loop is the existence of charged domain walls decorating the periphery of the repolarization nucleus. These walls oppose the growth of the switched domain and favor back-switching, thus creating a scenario of controlled volatile ferroelectric switching. Although the measurements were carried out with single crystals, prospectively volatile polarization switching can be tuned by tailoring sample thickness, domain wall mobility and electric fields, paving way to non-linear dielectric properties for smart electronic circuits.*

**INTRODUCTION**

The quest towards smart electronic circuits that emulate biological learning is motivated by artificial intelligence applications combined with the ever present need for lower energy consumption.[1-3] To achieve electronic building blocks of such circuits, materials that show neuromorphic-like functional properties (similar to neurons and synapses) have become of great interest. By now, many binary and complex oxides, chalcogenides, and 2D materials have been reported to exhibit memristive properties (e.g. non-volatile switchable electrical resistance) and several materials, primarily with metal-insulator transitions, also display neuristive properties where switching is volatile and the switched material quickly relaxes without applied bias.[4-10] Often, the switching is associated with electrically-driven breakdown phenomena, including filamentary breakdown and electro-thermal melting of the insulating state. Although breakdown produces sought-after non-linearities in the current/voltage (I/V) curves, it is also the major source of large energy cost of present neuromorphic devices and thus drives the search for new materials and/or approaches to implement neuromorphic circuitry.

Ferroelectric materials with non-volatile and electrically switchable polarization states have long been considered for high-efficiency memory applications in traditional CMOS electronics.[11, 12] Their much more recent connection to the realm of neuromorphics was enabled by the discovery of fundamental properties that couple electronic and polarization properties, enabling a path for electronic memory effects based on the domain configuration[7, 13, 14]. In particular, charged domain walls exhibit tunable and even metallic conductivity and are central to neuromorphic applications.[15-17] In general, robust control over ferroelectric switching is required for on-demand injection of charged domain walls, which are generally metastable. Meanwhile, ferroelectrics can in principle enable memcapacitive circuitry. Indeed, ferroelectric materials and their derivatives have long been known and are widely used in tunable dielectric applications. The challenge of memcapacitance, however, is that capacitance must exhibit a well-defined history of applied bias (or current), perhaps with at least two distinct non-volatile values. By contrast, tunable dielectric circuits are intentionally designed to exhibit little or no hysteresis. Here we will show that volatile ferroelectric switching is one path to enable memcapacitive phenomena without introducing atomic-scale disorder.

In this paper, we explore ferroelectric switching under conditions where nucleating domains will not achieve equilibrium geometry of their domain walls. In particular, we have probed localized ferroelectric switching in single crystals of uniaxial $Sn_2P_2S_6$[18-22] which is a lead and oxygen-free ferroelectric with a fairly small spontaneous polarization[23] of ~ 15 $\mu C/cm^2$ (compared to common oxide ferroelectrics with 3-10 fold larger polarization). The thickness of the crystal (~2 mm) was chosen so that ferroelectric switching by so-called domain breakdown[24] would not occur on the time-scale of the measurement. Our main research thrusts addressed 1) whether such volatile ferroelectric switching can significantly affect the shape of the ferroelectric hysteresis loop and 2) whether a combined effect of charged domain walls and non-equilibrium domain shapes can produce interesting electronic and dielectric signatures for neuromorphic concepts. We use near-field microwave scanning probes to locally modify and test electrical material properties. Remarkably, we find that the ferroelectric hysteresis loop is strongly controlled by the direction of the initial spontaneous polarization, where the loop shifted along the voltage axis towards the polarity that does not switch the initial polarization. Meanwhile, symmetric properties are observed near existing ferroelectric domain walls, ruling out a spectrum of external artefacts as a genesis of the observed phenomena. Given the large degree of polarization-induced shift, the regions of maximum dielectric tunability and microwave conductance both approach lower voltage thresholds. Therefore, local dielectric properties and conductivity in $Sn_2P_2S_6$ become dependent on the electric field history associated with domain nucleation and growth, enabling further consideration of $Sn_2P_2S_6$ a suitable candidate for memelectronic materials towards neuromorphic systems. Our results are supported by finite-elements numerical modeling of domain nucleation within Landau-Ginzburg -Devonshire theory and point to instabilities of newly switched domains in thick samples due to charged domain walls as the origin of the strong dependence of switching hysteresis on polarization orientation. Finally, we remark that although volatile switching may be undesirable for memory applications, it is a sought-after property for active circuits, such as relaxation oscillators, which can mimic neuronal spiking activity and exhibit a broad array of intriguing temporal dynamics.

## RESULTS AND DISCUSSION

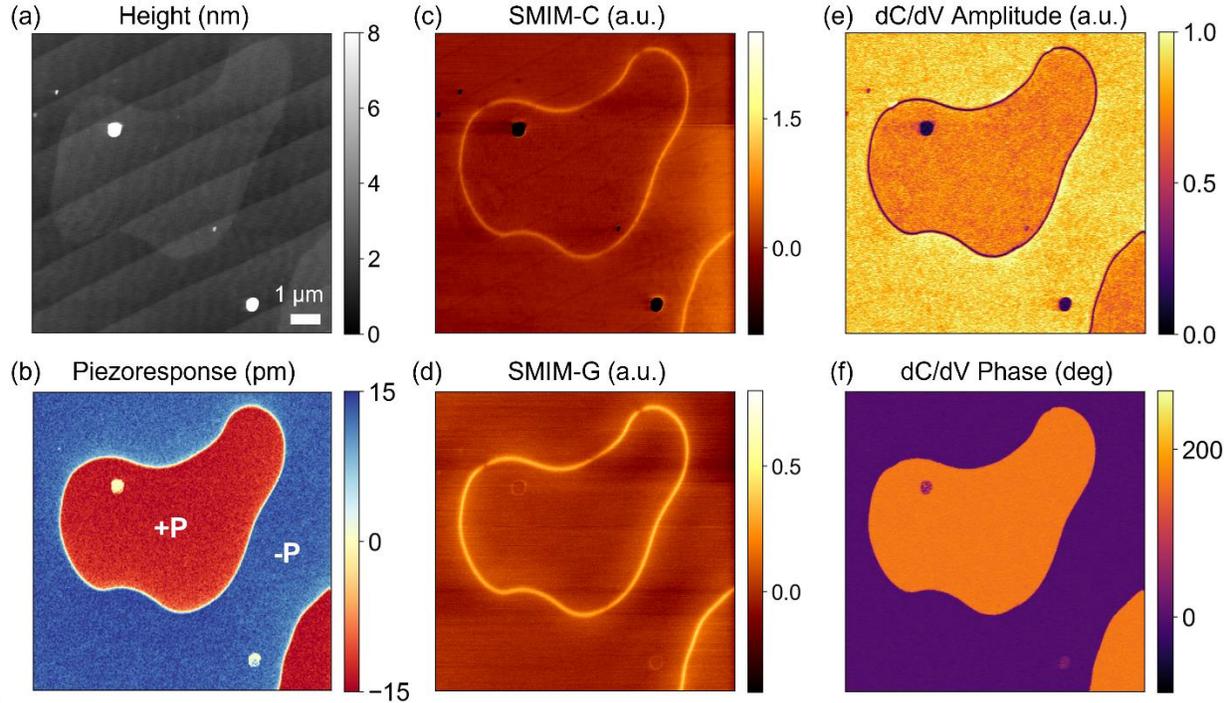

**Figure 1.** (a) Height, (b) PFM, (c) SMIM-C, (d) SMIM-G, (e) dC/dV amplitude and (f) dC/dV phase images of ferroelectric $Sn_2P_2S_6$ domains.

We use combined scanning microwave impedance microscopy (SMIM) and piezoresponse force microscopy (PFM) to probe ferroelectric, dielectric and conductive properties with nanoscale resolution.[13, 25-29] PFM measurements are carried out using band excitation around the contact resonance. Simultaneous microwave measurements are performed by transmitting with an AC electric field at ~3 GHz to the sample and measuring the microwaves reflected back into the probe and demodulation electronics. This reflected microwave signal is demodulated into two analog signals corresponding to the real and imaginary parts of the reflected wave. The real part corresponds to the resistivity or, inversely, the local conductivity $G$ of the sample based on how much of the AC excitation field dissipated into the sample. The imaginary part is associated with the capacitance $C$ of the sample-probe system which is a function of both, the sample dielectric constant $\varepsilon$ and the capacitor geometry $C = \varepsilon A/d$ (with the effective electrode area $A$ and the distance *d*). The SMIM-C signal can therefore be sensitive to dielectric properties and sample topography alike.

The topography (Figure 1(a)) presents as flat surface with steps visible on the cleaved surface and a slightly elevated topography at the domain where the polarization vector is oriented along the +z direction and therefore corresponds to a +P domain, as labeled in the PFM image (Figure 1(b)). The surrounding area mostly shows a -P polarization, which means the polarization vector is oriented along the -z direction. Note that the domain walls (DWs) do not fully coincide with this topographic feature. Moreover, cleaving did not occur strictly along a crystallographic axis, and therefore a small in-plane polarization may be present. SMIM-C images reveal a higher signal at domain walls which is associated with a higher dielectric constant, compared to the domain surfaces. As commonly observed, dust particles appear as low signals in the SMIM-C channel due to the topographic crosstalk and dielectric properties as discussed previously. This topographic crosstalk due to the SMIM-C signal formation mechanism is also evident from the contrast associated with steps. The SMIM-G signal, however, is free from topographic crosstalk like the steps visible in the SMIM-C channel. The measured SMIM-G signal at the ferroelectric DWs is higher than at the domain surface and has a halo-like appearance in the vicinity of DWs.

In general, the contrast in SMIM-G signals at DWs has been linked to different phenomena:[27] (i) AC DW conductivity of charge carriers, as observed in $Pb(Zr_{0.2}Ti_{0.8})O_3$ and $KNbO_3$ [27, 28] and (ii) Dielectric losses that occur due to domain wall vibrations as the oscillation of bound charge is equivalent to mobile-carrier conduction in a microwave circuit (e.g. on $BiFeO_3$ or h-$RMnO_3$).[26, 30] In order to elucidate the likely origin of the enhanced SMIM-G signal at DWs in $Sn_2P_2S_6$, we extracted the change in SMIM-G across domain walls in dark and bright conditions (i.e. where the light contained a green/yellowish wavelength corresponding to the band gap of 2.2 eV).[18] As shown in the Supporting Information Figure S1, the conductivity increases in general under illumination as evident from a large offset. However, the relative change in SMIM-G signal across DWs and domain surfaces remains similar. This observation indicates that the enhanced losses at DWs compared to domain surfaces is not subject to availability of free charge carriers and points to DW oscillations as the origin of the signal. In addition, ferroelectrics such as $Pb(Zr_{0.2}Ti_{0.8})O_3$ (where SMIM-G shows a higher conductivity at DWs due to charge carriers) do not exhibit any DW contrast in the SMIM-C channel.[28] On the other hand, $BiFeO_3$ [26] and hexagonal rare-earth manganites[30] do show contrast in SMIM-C, consistent with our observations.

The variations of the capacitance in response to an AC voltage modulation at 15kHz is shown in Figure (1(e,g)). While the dC/dV amplitude indicates the magnitude of dielectric tunability, the phase of the response signal is associated with the polarity of the bound charge on the surface. The dC/dV amplitude signal slightly varies between domain surfaces and decreases to zero at domain walls, and the dC/dV phase coincides with ferroelectric domain polarities. Therefore, the dC/dV phase signal can be used to infer the orientation of the ferroelectric polarization.

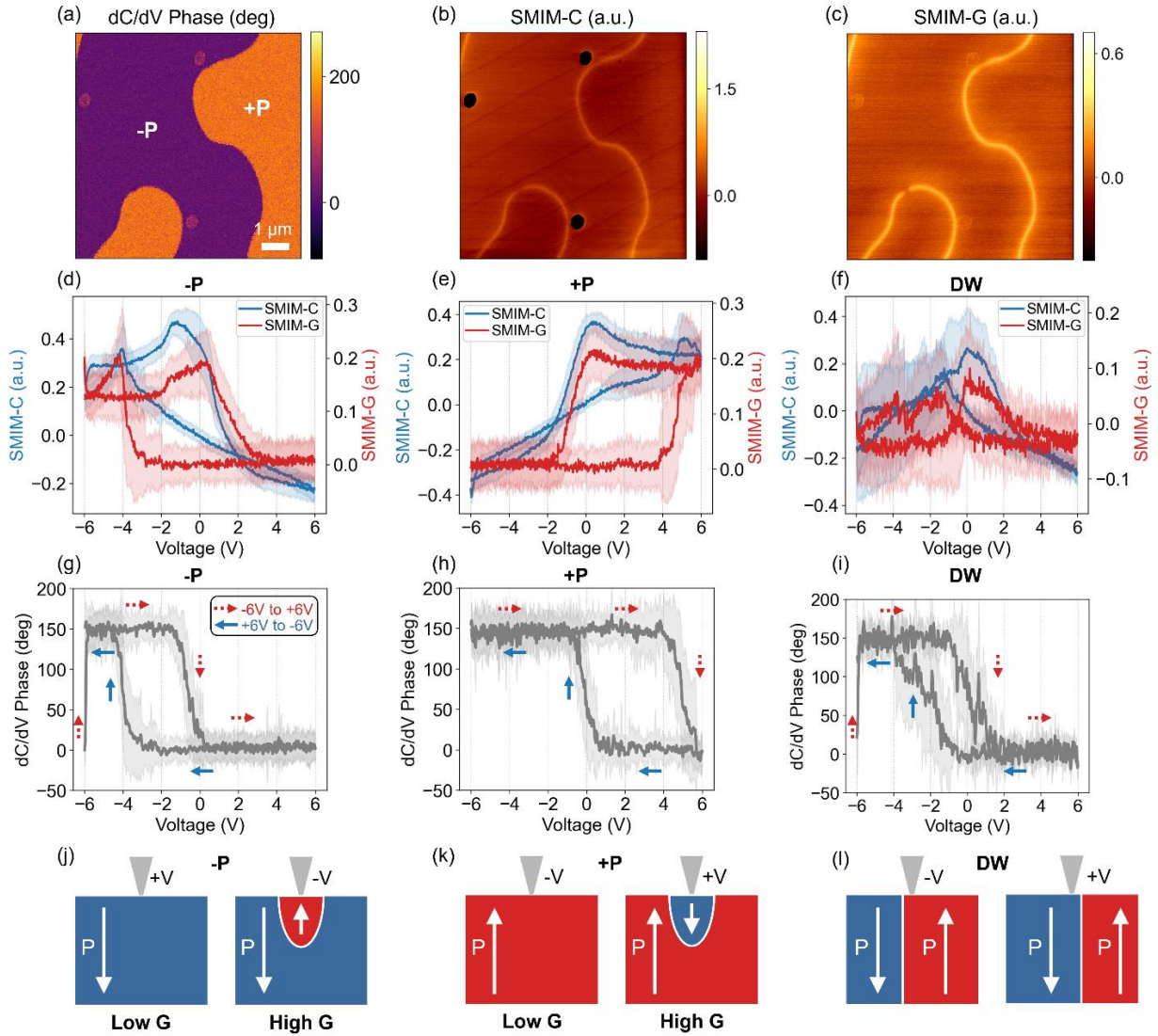

**Figure 2.** (a) dC/dV phase, (b) SMIM-C and (c) SMIM-G signals of $Sn_2P_2S_6$ domains where VS was performed. Mean (solid line) and standard deviation (shaded areas) SMIM-C and SMIM-G loops as functions of the applied voltage for (d) -P and (e) +P domain surfaces as well as (f) DW. dC/dV phase loops for (g) initial -P surface, (h) +P surface and (i) at DW. The arrows in (g,h,i) indicate the voltage sweep

direction (dashed red: -6 V to +6 V, solid blue: +6 V to -6 V). Schematic plots of ferroelectric domains during VS for areas with initial (j) -P and (k) +P polarization orientation and (l) domain walls. White arrows in (j,k,l) indicate polarization vectors.

To dynamically probe DW response, we apply voltage spectroscopy (VS) as this technique typically leads to the creation and movement of DWs as a function of the applied voltage. A DC voltage sweep from -6 V to +6 V and back to -6 V was performed at a rate of 30.5 V/s on the bulk $Sn_2P_2S_6$ crystal of ~2 mm thickness. VS was conducted as a grid of 8×8 pixels across the area depicted in Figure 2. As evident from the dC/dV phase image in Figure 2(a), the area covers domains of -P and +P orientation, separated by DWs prominently visible in SMIM-C and SMIM-G images as higher signal compared to domain surfaces (Figure 2(b,c), respectively). The domain structure is the same before and after VS as evident from the dC/dV phase images in the Supporting Information Figure S2(a,b). The absence of permanent switching is expected for the chosen experimental parameters where the sample thickness of ~2 mm combined with only a few volts applied to the tip results in electric fields that are not high enough to fully propagate domain walls through the crystals and form stable domains. However, due to the high electric field strength around scanning probe microscopy tips, it is possible to nucleate and grow temporary new domains and therefore create and move domain walls as a function of voltage. The SMIM-C, SMIM-G loops as a function of DC voltage are shown as for pixels located at -P or +P domain surfaces and DWs (Figure 2(d,e,f), respectively). Temporary domain nucleation and growth are indicated by dC/dV phase loops that were simultaneously recorded with SMIM-C and SMIM-G signals during VS (Figure 2(g,h,i)). For SMIM-C, SMIM-G and dC/dV phase loops shown in the same plot for of +P, -P and DW areas, see the Supporting Information, Figure S3.

The SMIM-C and SMIM-G loops show distinctively different behavior dependent on whether a pixel is located within the domain of initial -P or +P polarization or located at a DW. For the -P domain surface, the SMIM-C capacitance starts out fairly constant, before a steep decrease is observed around low negative voltages (Figure 2(d)). The reverse sweep from +6 V to -6 V results in a near linear increase in SMIM-C signal until previous levels are reached. Simultaneously, SMIM-G behaves fully hysteretically with a higher conductivity state at negative voltages. The dC/dV phase loop indicates that at the local polarization under the tip switches from -P to +P immediately after starting the sweep at -6 V, which is expected for a ferroelectric material (Figure 2(g)). As the voltage reaches zero, the polarization switches back to -P until voltages higher than -4 V are applied. The +P domain exhibits near-linear increase in SMIM-C capacitance during the voltage sweep from -6 V to +6 V (Figure 2(e)). In the reverse sweep direction, a plateau is evident followed by a steep decrease starting at 0 V. SMIM-G shows hysteretic behavior with higher conductivity at positive voltages, which according to the dC/dV phase data (Figure 2(h)) are associated with the nucleation of a -P domain. For both domain polarizations, the more constant SMIM-C state corresponds to the higher SMIM-G state. The strong imprint of dC/dV phase loops towards negative (for initial -P domains) and positive (for initial +P domains) voltages corroborates that any domain nucleation and growth is not stable and readily relaxes back to the initial domain polarization at voltages close to 0 V. At DWs, SMIM-C and -G loops are more reminiscent of ferroelectric switching of materials with conducting domain walls (Figure 2(f)) and the dC/dV phase loop is less imprinted (Figure 2(i)).[28]

The higher SMIM-G values correspond to polarization switching events, as the tip interacts with domain walls. These observations indicate that DWs can be moved sideways back and forth if the tip is placed nearby, as opposed to nucleating a new domain within a domain of opposite polarization. Figures 2(j,k,l) schematically summarize and display ferroelectric behavior for pixels located at initial -P, +P and DWs, respectively. Higher SMIM-G is linked to the creation of unstable half-prolate shaped domains. As polarization vectors are oppositely oriented between the main and bubble domain, the DWs between them are charged and unstable, and therefore likely to oscillate in response to the AC electric field. Moreover, due to the 3-dimensional shape of the bubble domain, the size of the DW interface is expected to be large within the SMIM probing volume, leading to larger dissipation of the microwave signal compared to sideways DW motion. Given the similar values for SMIM-G, it appears that domain growth is similar independent of whether -P forms within a +P matrix or vice versa.

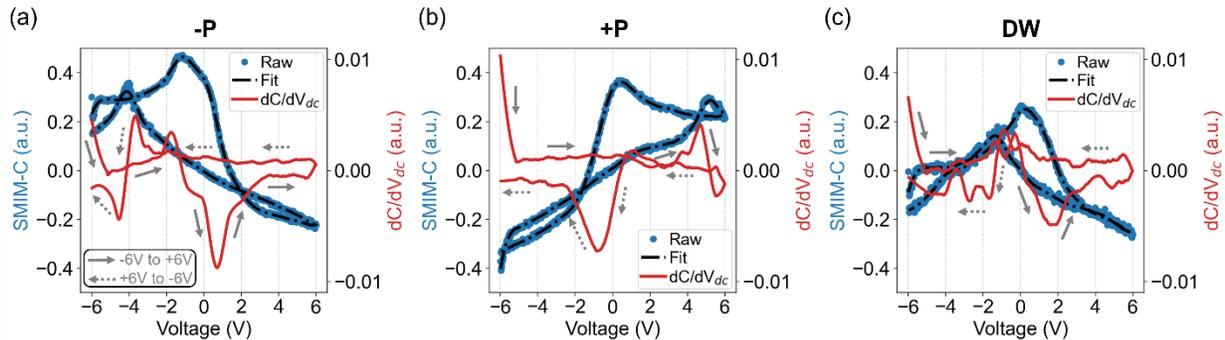

**Figure 3.** Raw SMIM-C signal (blue dots), fitted SMIM-C signal (black dashed line) and derivative dC/dV$_{dc}$ (red solid line) for (a) -P and (b) +P domain areas and at (c) DWs. Arrows indicate dC/dV$_{dc}$ loop direction from -6 V to + 6 V (solid) and +6 V to -6 V (dashed).

Tunability of dielectric properties in response to a 15 kHz AC electric field is evident from the strong dC/dV$_{ac}$ signal in Figure 1(e,f). To analyze tunability in response to DC electric fields, we calculate dC/dV$_{dc}$ from the fitted SMIM-C loops as plotted in Figure 3. Significant changes in SMIM-C as a function of DC voltage coincide with switching events for +P and -P domain surfaces as well as DWs (see Figure 3(a), (b), and (c), respectively). The dielectric properties change strongest when the nucleated domain is switched back to the initial polarization for -P and +P alike. In comparison, moving existing DWs results in a less pronounced change in SMIM-C. Therefore, SMIM-G and SMIM-C can be simultaneously controlled through reversible nucleation of domains. We further note that $Sn_2P_2S_6$ evinces a relatively low Curie temperature of ~80° C and can be lowered even further through doping or strain effects;[31] this tunability of structure-property relationships in this material also makes it relevant for industrial applications underpinned by high-frequency tunability.

To obtain further insight in the polarization reversal and domain growth process in $Sn_2P_2S_6$ and understand the strong asymmetry of the hysteresis loops measured by SMIM, we performed finite-elements numerical simulations using a Landau-Ginzburg-Devonshire (LGD) approach. Details of the LGD approach applied for simulation of polarization reversal in ferroelectrics can be found in Refs.[32, 33] The simulations were carried out in cylindrically symmetrical geometry with the axis of symmetry passed though the center of the tip. Polarization switching was induced by the electric field of the biased AFM tip with circular contact area with radius a of *$R_{cont}$ = 100 nm* and a tip apex curvature of *$R_{tip}$ = 100 nm*. Two sample geometries were studied: (1) a thin film with a thickness *$h_{thin}$ = 100 nm* and (2) a thick film with a thickness *$h_{thick}$ = 4000 nm*. In both cases the diagonal tensor of dielectric permittivity with *$\varepsilon_{11} = \varepsilon_{22} = 84$ and $\varepsilon_{33} = 30$* was used. We used values of spontaneous polarization *$P_s = 3$ μC/cm$^2$* and coercive field *$E_c = 10$ kV/mm* to calculate the LGD parameters (*α* and *β*). A single rectangular pulse of the electric bias with amplitude of 10 V and duration ranging from 250 ms to 750 ms was applied to the tip in the simulations. After termination of the pulse, the bias was dropped to 0 V to study relaxation of the switched structure for up to 10 s.

Resulting maps illustrating the polarization evolution at different time steps and values of applied bias are shown in figure 4. As one can see, the switching process differs significantly for thin and thick samples. In the case of the thick sample (Fig. 4a,b), the domain nucleates underneath the tip and quickly forms a conical shape, which propagates toward the bottom electrode. After the pulse termination, backswitching occurs underneath the tip, which leads to complete disappearance of the formed domain in the case of a short (250 ms) switching pulse (Fig. 4a). In the case of a long pulse, backswitching leads to formation of the residual domain structure, which further propagates down to the bottom electrode under the action of breakthrough / brakedown phenomena, which are caused by the configuration of electric fields in the vicinity to the tip of the domain and can lead to domain growth far from the areas with externally applied fields.[24, 34]

In the case of the thin film (Fig. 4c,d), the newly formed domain quickly reaches the bottom electrode and further growth mostly occurs due to the lateral motion of vertical domain walls. In contrary to the case of the thick film, the resulting structures are stable and do not undergo any significant changes after termination of the switching bias.

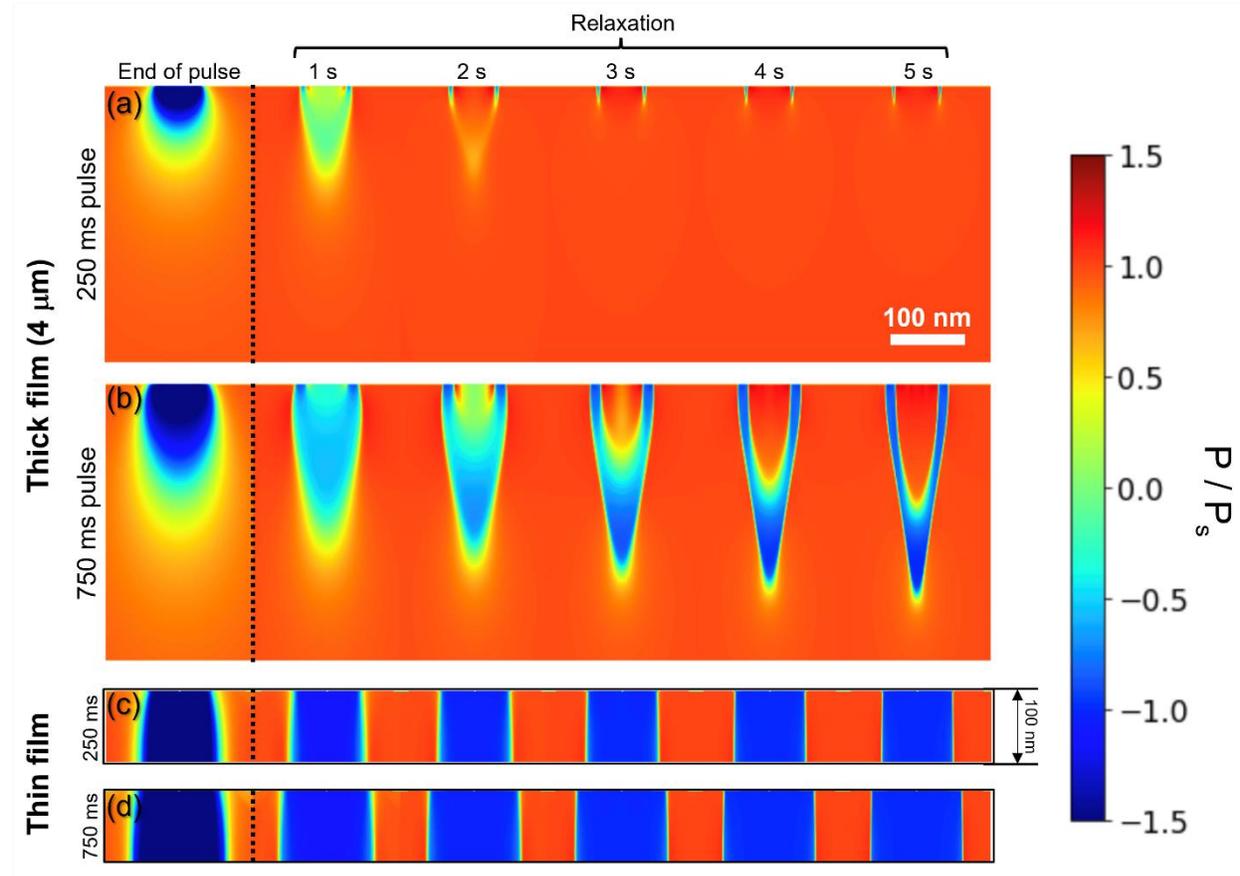

**Figure 4.** Finite-elements numerical simulations of the polarization switching that occurs in a thin film and a thick sample as a function of the duration of a single rectangular bias pulse. Cross sectional maps showing the distribution of switched regions for (a,b) the thin film and (c,d) the thick sample after application of switching pulses with a duration of 250 ms (a,c) and 750 ms (b,d). The red and blue colors correspond to positive and negative polarization, respectively, as indicated in the color scale bar.

Comparison of the two considered sample configurations shows dramatic differences in the process of polarization reversal and backswitching for thin and thick samples. Switching in a thin film is mostly controlled by the distribution of the externally applied electric field and occurs due to the lateral motion of uncharged vertical domain walls. This leads to a much more stable behavior than in the case of a thick sample, where the presence of charged domain walls significantly contribute to unstable domain structures. To better highlight the different types of switching behavior, we plotted dependences of the switched volume (areas with $P / P_s < 0$ mC/cm$^2$) as functions of the time in all considered cases (Fig. 5). It can be clearly seen that in the case of a thick sample, the original domain polarization is preferred due to domain instabilities arising from charged subsurface domain walls, which explains the strong imprint of the loops observed in the experimental data.

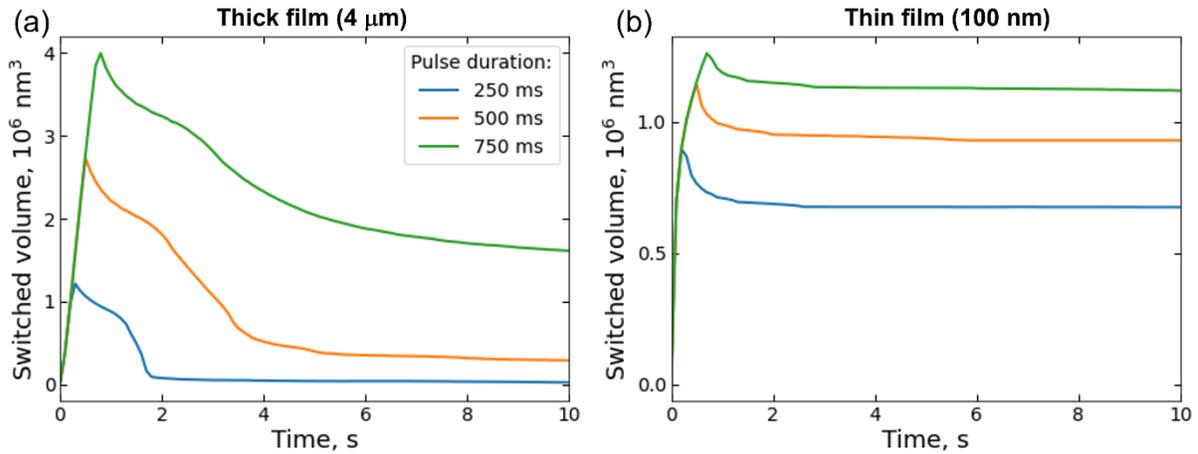

**Figure 5.** Simulated plots of the switched volume as function of time for (a) a thin film and (b) a thick sample.

The balance between volatile and non-volatile switching in our case is therefore a natural consequence of the experiment geometry, where the switched volume is significantly smaller than the thickness and the width of the ferroelectric material. To achieve memcapacitive phenomena through controlled imprint, one can generally pursue other approaches such as dielectric spacer layers, controlled disorder, or even in-plane switching in multiaxial geometry. We hypothesize that direct measurement of the capacitive properties of such geometries will provide a better understanding of the ferroelectric dynamics and also increase the spectrum of devices where memcapacitive phenomena can be implemented and controlled for more complex neuromorphic functions.

**CONCLUSIONS**

We have demonstrated a nanoscale tunable and polarization dependent behavior of ferroelectric $Sn_2P_2S_6$ that is underpinned by reversible nucleation and instabilities of domains. Both dielectric properties and conductivity are dependent on voltage history and are able to be modified using the nucleation, growth, and oscillation of DWs as a tuning knob. We envision memelectronic devices based on transient ferroelectric switching to create a specific domain matrix to tailor subsequent resistive and capacitive response to low voltages dependent on whether domain nucleation occurs. Moreover, the origin of the enhanced SMIM-G signal was identified as domain wall vibrations, consistent with literature[26-28] and behavior under illumination. We envision these findings to provide a path towards neuromorphic nanoelectronics based on tunable memristive and memcapacitive structures.

**MATERIALS AND METHODS**

Single crystals of $Sn_2P_2S_6$ were synthesized via the vapor transport technique. Briefly, elemental Sn (Alfa Aesar Puratronic 99.9999+%), P (Alfa Aesar Puratronic 99.999+%) and S (Alfa Aesar Puratronic 99.9995%) in the near-stoichiometric ratio Sn:P:S 2:2.2:6. Elements were sealed in a 12 cm long ampoule with a 13 mm ID together with ~100-200 mg $I_2$ (Alfa Aesar, 99.5%) as a vapor transport agent. The sample was placed in a two-zone reflecting furnace with the hot end at 700ºC and the cold end at 650ºC. The crystals were allowed to nucleate and grow for 100 h, yielding naturally faceted samples with a maximum diameter of 15 mm. We used X-ray diffraction via a Bruker D8 Discover Davinci with Cu K$\alpha$ radiation to determine the orientation of the facets and polished the crystals accordingly. The $Sn_2P_2S_6$ crystal was mechanically cleaved and a sample of ~2 mm thickness was chosen for measurements.

SMIM measurements were performed in ambient conditions on a Bruker Icon instrument. The PrimeNano Scanwave™ Pro electronics module was used to generate microwaves and detect the reflected signal. Coaxially shielded probes (nominal force constant 8 N/m, nominal resonance frequency 75 kHz, tip material TiW) were used. To separate SMIM-C from SMIM-G signals, the demodulation angle was adjusted until only capacitive contrast was measured on an Al dot standard sample with metal electrodes on a $SiO_2$ substrate. For improved separation, SMIM signals were further separated by post-processing in Python using vector rotation using vector rotation of the combined SMIM-C and SMIM-G signals treated as vector components with minimization of the signal from the dust particles in the SMIM-G channel as the optimization criterium.

Finite element LGD simulations of the domain growth for a thick and a thin film was performed using COMSOL Multiphysics software suite. More details about application of LGD approach for simulation of ferroelectric switching can be found in the following papers. [32, 33]

## ACKNOWLEDGEMENTS

The experiments and simulations were supported by the Center for Nanophase Material Sciences, which is a U.S. DOE Office of Science User Facility. Analysis and manuscript writing were supported by the U.S. Department of Energy (DOE), Office of Science, Basic Energy Sciences, Materials Sciences and Engineering Division. We also acknowledge support through the United States Air Force Office of Scientific Research (AFOSR) LRIR 18RQCOR100 as well as funding by AOARD-MOST Grant Number F4GGA21207H002. In addition, we acknowledge funding from the National Academies of Science and Engineering through the National Research Council Senior Fellowship award. A.T. is grateful to the FCT, Portugal, for the CEEC 2021 Individual support as well as acknowledges support of the project CICECO-Aveiro Institute of Materials, UIDB/50011/2020 & UIDP/50011/2020, financed by national funds through the FCT/MEC and when appropriate co-financed by FEDER under the PT2020 Partnership Agreement.

# REFERENCES

1. Boni, G. A.; Filip, L. D.; Chirila, C.; Iuga, A.; Pasuk, I.; Hrib, L.; Trupina, L.; Pintilie, I.; Pintilie, L., Memcomputing and Nondestructive Reading in Functional Ferroelectric Heterostructures. *Physical Review Applied* **2019,** *12* (2), 024053.
2. del Valle, J.; Salev, P.; Kalcheim, Y.; Schuller, I. K., A caloritronics-based Mott neuristor. *Scientific Reports* **2020,** *10* (1), 4292.
3. Pickett, M. D.; Medeiros-Ribeiro, G.; Williams, R. S., A scalable neuristor built with Mott memristors. *Nature Materials* **2013,** *12* (2), 114-117.
4. Sarwat, S. G.; Moraitis, T.; Wright, C. D.; Bhaskaran, H., Chalcogenide optomemristors for multi-factor neuromorphic computation. *Nature Communications* **2022,** *13* (1), 2247.
5. Peng, Z.; Wu, F.; Jiang, L.; Cao, G.; Jiang, B.; Cheng, G.; Ke, S.; Chang, K.-C.; Li, L.; Ye, C., HfO2-Based Memristor as an Artificial Synapse for Neuromorphic Computing with Tri-Layer HfO2/BiFeO3/HfO2 Design. *Advanced Functional Materials* **2021,** *31* (48), 2107131.
6. Li, Y.; Zhong, Y.; Xu, L.; Zhang, J.; Xu, X.; Sun, H.; Miao, X., Ultrafast Synaptic Events in a Chalcogenide Memristor. *Scientific Reports* **2013,** *3* (1), 1619.
7. Chanthbouala, A.; Garcia, V.; Cherifi, R. O.; Bouzehouane, K.; Fusil, S.; Moya, X.; Xavier, S.; Yamada, H.; Deranlot, C.; Mathur, N. D.; Bibes, M.; Barthélémy, A.; Grollier, J., A ferroelectric memristor. *Nature Materials* **2012,** *11* (10), 860-864.
8. Cao, G.; Meng, P.; Chen, J.; Liu, H.; Bian, R.; Zhu, C.; Liu, F.; Liu, Z., 2D Material Based Synaptic Devices for Neuromorphic Computing. *Advanced Functional Materials* **2021,** *31* (4), 2005443.
9. Ilyas, N.; Li, D.; Li, C.; Jiang, X.; Jiang, Y.; Li, W., Analog Switching and Artificial Synaptic Behavior of Ag/SiOx:Ag/TiOx/p++-Si Memristor Device. *Nanoscale Research Letters* **2020,** *15* (1), 30.
10. Ma, Y.; Yeoh, P. P.; Shen, L.; Goodwill, J. M.; Bain, J. A.; Skowronski, M., Evolution of the conductive filament with cycling in TaOx-based resistive switching devices. *Journal of Applied Physics* **2020,** *128* (19), 194501.
11. Scott, J. F., *Ferroelectric Memories*. 1 ed.; Springer-Verlag Berlin, Heidelberg: 2000.
12. Mikolajick, T.; Schroeder, U.; Slesazeck, S., The Past, the Present, and the Future of Ferroelectric Memories. *IEEE Transactions on Electron Devices* **2020,** *67* (4), 1434-1443.
13. Burns, S. R.; Tselev, A.; Ievlev, A. V.; Agar, J. C.; Martin, L. W.; Kalinin, S. V.; Sando, D.; Maksymovych, P., Tunable Microwave Conductance of Nanodomains in Ferroelectric PbZr0.2Ti0.8O3 Thin Film. *Advanced Electronic Materials* **2021,** *n/a* (n/a), 2100952.
14. Kim, D. J.; Lu, H.; Ryu, S.; Bark, C. W.; Eom, C. B.; Tsymbal, E. Y.; Gruverman, A., Ferroelectric Tunnel Memristor. *Nano Letters* **2012,** *12* (11), 5697-5702.
15. Lu, H.; Tan, Y.; McConville, J. P. V.; Ahmadi, Z.; Wang, B.; Conroy, M.; Moore, K.; Bangert, U.; Shield, J. E.; Chen, L.-Q.; Gregg, J. M.; Gruverman, A., Electrical Tunability of Domain Wall Conductivity in LiNbO3 Thin Films. *Advanced Materials* **2019,** *31* (48), 1902890.
16. Sharma, P.; Moise, T. S.; Colombo, L.; Seidel, J., Roadmap for Ferroelectric Domain Wall Nanoelectronics. *Advanced Functional Materials* **2022,** *32* (10), 2110263.
17. Wang, C.; Wang, T.; Zhang, W.; Jiang, J.; Chen, L.; Jiang, A., Analog ferroelectric domain-wall memories and synaptic devices integrated with Si substrates. *Nano Research* **2022,** *15* (4), 3606-3613.
18. Li, Y.; Singh, D. J., Properties of the ferroelectric visible light absorbing semiconductors: Sn2P2S6 and Sn2P2Se6. *Physical Review Materials* **2017,** *1* (7), 075402.

19.	Studenyak, I. P.; Mitrovcij, V. V.; Kovacs, G. S.; Mykajlo, O. A.; Gurzan, M. I.; Vysochanskii, Y. M., Temperature variation of optical absorption edge in Sn2P2S6 and SnP2S6 crystals. *Ferroelectrics* **2001,** *254* (1), 295-310.
20.	Yu, M. V.; Yu, M. V.; Yu, M. V.; Yevych, R. M.; Yevych, R. M.; Yevych, R. M., The second order phase transition in Sn2P2S6 crystals: anharmonic oscillator model. *Condensed Matter Physics* **2008,** *11* (3), 417-417.
21.	Glukhov, K.; Fedyo, K.; Banys, J.; Vysochanskii, Y., Electronic structure and phase transition in ferroelectic Sn(2)P(2)S(6) crystal. *Int J Mol Sci* **2012,** *13* (11), 14356-14384.
22.	Yevych, R. M.; Vysochanskii, Y. M., Triple Well Potential and Macroscopic Properties of Sn2P2S6 Ferroelectrics Near Phase Transition. *Ferroelectrics* **2011,** *412* (1), 38-44.
23.	Zamaraite, I.; Yevych, R.; Dziaugys, A.; Molnar, A.; Banys, J.; Svirskas, S.; Vysochanskii, Y., Double Hysteresis Loops in Proper Uniaxial Ferroelectrics. *Physical Review Applied* **2018,** *10* (3), 034017.
24.	Molotskii, M.; Agronin, A.; Urenski, P.; Shvebelman, M.; Rosenman, G.; Rosenwaks, Y., Ferroelectric Domain Breakdown. *Physical Review Letters* **2003,** *90* (10), 107601.
25.	Chu, Z.; Zheng, L.; Lai, K., Microwave Microscopy and Its Applications. *Annual Review of Materials Research* **2020,** *50* (1), 105-130.
26.	Huang, Y.-L.; Zheng, L.; Chen, P.; Cheng, X.; Hsu, S.-L.; Yang, T.; Wu, X.; Ponet, L.; Ramesh, R.; Chen, L.-Q.; Artyukhin, S.; Chu, Y.-H.; Lai, K., Unexpected Giant Microwave Conductivity in a Nominally Silent BiFeO3 Domain Wall. *Advanced Materials* **2020,** *32* (9), 1905132.
27.	Schultheiß, J.; Rojac, T.; Meier, D., Unveiling Alternating Current Electronic Properties at Ferroelectric Domain Walls. *Advanced Electronic Materials* **2021,** *n/a* (n/a), 2100996.
28.	Tselev, A.; Yu, P.; Cao, Y.; Dedon, L. R.; Martin, L. W.; Kalinin, S. V.; Maksymovych, P., Microwave a.c. conductivity of domain walls in ferroelectric thin films. *Nature Communications* **2016,** *7* (1), 11630.
29.	Wu, H.; Zhou, J.; Lan, C.; Guo, Y.; Bi, K., Microwave Memristive-like Nonlinearity in a Dielectric Metamaterial. *Scientific Reports* **2014,** *4* (1), 5499.
30.	Wu, X.; Petralanda, U.; Zheng, L.; Ren, Y.; Hu, R.; Cheong, S.-W.; Artyukhin, S.; Lai, K., Low-energy structural dynamics of ferroelectric domain walls in hexagonal rare-earth manganites. *Science Advances* **2017,** *3* (5), e1602371.
31.	Rizak, I. M.; Rizak, V. M.; Vysochansky, Y. M.; Gurzan, M. I.; Slivka, V. Y., Tricritical Lifshitz point in phase diagram of (PbySn1-y)2P2(SexS1-x)6 ferroelectrics. *Ferroelectrics* **1993,** *143* (1), 135-141.
32.	Morozovska, A. N.; Ievlev, A. V.; Obukhovskii, V. V.; Fomichov, Y.; Varenyk, O. V.; Shur, V. Y.; Kalinin, S. V.; Eliseev, E. A., Self-consistent theory of nanodomain formation on nonpolar surfaces of ferroelectrics. *Physical Review B* **2016,** *93* (16), 165439.
33.	Ievlev, A. V.; Morozovska, A. N.; Shur, V. Y.; Kalinin, S. V., Humidity effects on tip-induced polarization switching in lithium niobate. *Applied Physics Letters* **2014,** *104* (9), 092908.
34.	Morozovska, A. N.; Eliseev, E. A.; Li, Y.; Svechnikov, S. V.; Maksymovych, P.; Shur, V. Y.; Gopalan, V.; Chen, L.-Q.; Kalinin, S. V., Thermodynamics of nanodomain formation and breakdown in scanning probe microscopy: Landau-Ginzburg-Devonshire approach. *Physical Review B* **2009,** *80* (21), 214110.

# SUPPORTING INFORMATION

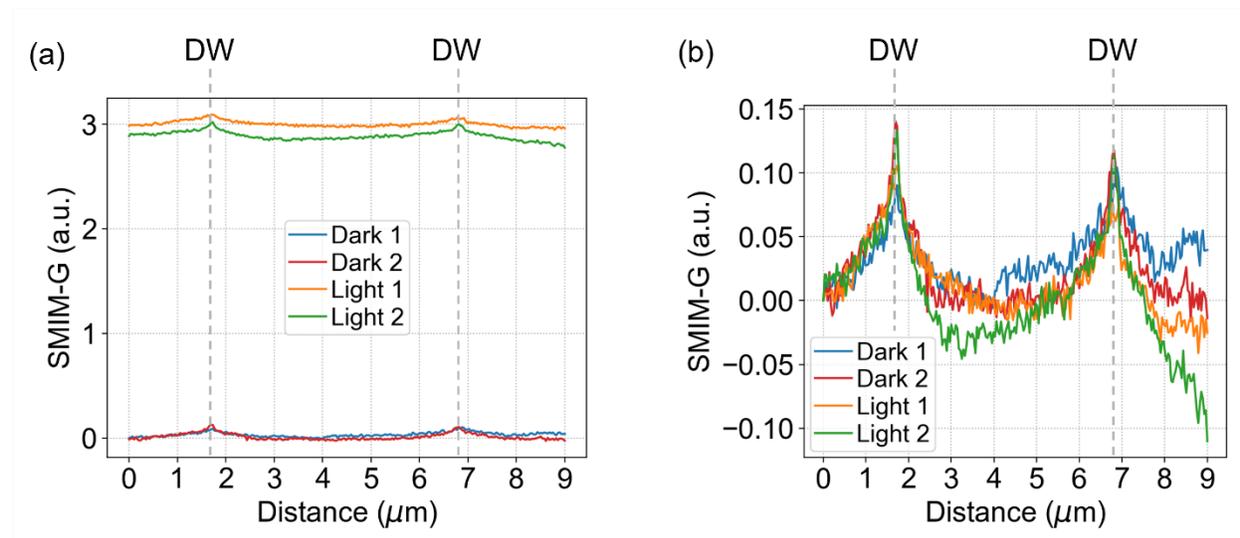

**Figure S1.** SMIM-G signal scanning across the same line that includes two domain walls at indicated positions with the slow scan axis disabled. (a) Measured SMIM-G data and (b) offset applied to start at 0 to compare the peaks at domain walls. The profiles are averaged over three lines.

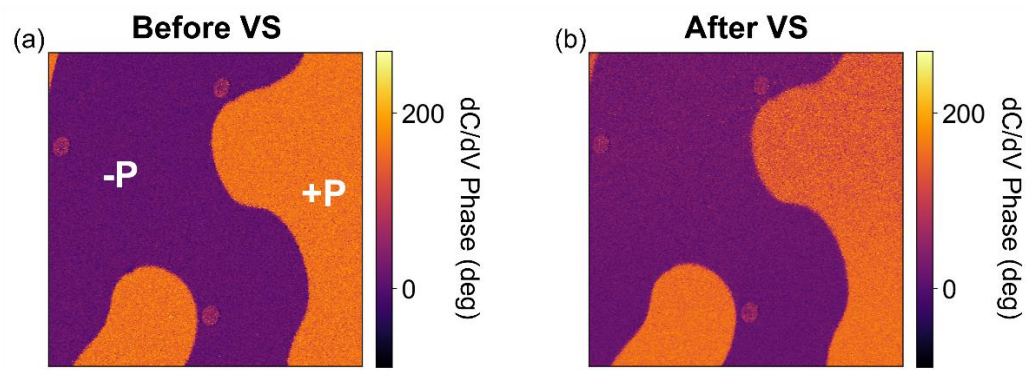

**Figure S2.** Height and dC/dV Phase images (a,b) before and (c,d) after voltage spectroscopy (VS). Height images indicate preferential particle growth on +P domain surfaces, dC/dV images show no permanent polarization switching.

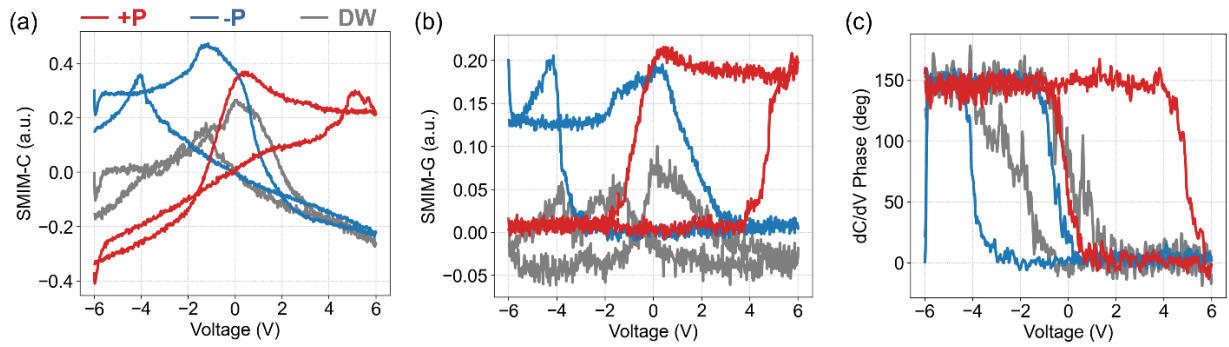

**Figure S3.** Loops of mean (a) SMIM-C, (b) SMIM-G, and (c) dC/dV phase signals as a function of DC voltage measured on +P, -P and DW.